\documentclass[conference]{IEEEtran}
\IEEEoverridecommandlockouts
\usepackage{cite}
\usepackage{amsmath,amssymb,amsfonts}
\usepackage{algorithmic}
\usepackage{graphicx}
\usepackage{textcomp}
\usepackage{xcolor}
\usepackage{braket}
\usepackage{hyperref}

\def\BibTeX{{\rm B\kern-.05em{\sc i\kern-.025em b}\kern-.08em
    T\kern-.1667em\lower.7ex\hbox{E}\kern-.125emX}}
\begin{document}

\makeatletter
\newcommand{\newlineauthors}{%
  \end{@IEEEauthorhalign}\hfill\mbox{}\par
  \mbox{}\hfill\begin{@IEEEauthorhalign}
}
\makeatother

\title{A Survey of Classical And Quantum Sequence Models}
\author{\IEEEauthorblockN{1\textsuperscript{st} I-Chi Chen }
\IEEEauthorblockA{\textit{ Department of Physics and Astronomy,}\\ \textit{ Iowa State University, Ames, IA, USA} \\
ichen@iastate.edu}
\and
\IEEEauthorblockN{2\textsuperscript{nd} Harshdeep Singh}
\IEEEauthorblockA{\textit{Centre for Computational and Data Sciences,}\\ \textit{IIT Kharagpur, India} \\
harshdeeps@kgpian.iitkgp.ac.in}
\newlineauthors
\IEEEauthorblockN{3\textsuperscript{rd} V L Anukruti} 
\IEEEauthorblockA{\textit{CQST}\\\textit{IIIT Hyderabad, India}\\anuvadali@proton.me}
\and
\IEEEauthorblockN{4\textsuperscript{th} Brian Quanz}
\IEEEauthorblockA{\textit{IBM Quantum} \\
\textit{IBM Research, USA}\\
blquanz@us.ibm.com}
\and
\IEEEauthorblockN{5\textsuperscript{th} Kavitha Yogaraj}
\IEEEauthorblockA{\textit{IBM Quantum} \\
\textit{IBM Research, India}\\
kyogarj1@in.ibm.com}
}






\maketitle
\IEEEpubidadjcol
\begin{abstract}
Our primary objective is to conduct a brief survey of various classical and quantum neural net sequence models, which includes self-attention and recurrent neural networks, with a focus on recent quantum approaches proposed to work with near-term quantum devices, while exploring some basic enhancements for these quantum models.  
We re-implement a key representative set of these existing methods, adapting an image classification approach using quantum self-attention to create a quantum hybrid transformer that works for text and image classification, and applying quantum self-attention and quantum recurrent neural networks to natural language processing tasks. We also explore different encoding techniques and introduce positional encoding into quantum self-attention neural networks leading to improved accuracy and faster convergence in text and image classification experiments. This paper also performs a comparative analysis of classical self-attention models and their quantum counterparts, helping shed light on the differences in these models and their performance. 

\end{abstract}


\section{Introduction}
Sequence models~\cite{h1} are machine learning models that take as input and/or output sequences of data. Sequential data includes time series, text, audio, video, etc. Over the past couple decades, neural networks have arguably become the most popular type of model used for sequence modeling across diverse applications.  Recurrent Neural Networks (RNNs)~\cite{h2} were one of the first neural network architectures designed for sequential data, and RNN models have become a key, popular class of algorithms used for sequence modeling. Transformer models~\cite{h3} were developed to bypass the sequential, recurrent computation required by RNNs, which can require many sequential steps and hidden state parameters to model relationships between two arbitrary positions in a sequence. Instead with Transformer models a single parallelizable step is used to update each sequence position based on all others,  using an attention-weighting approach.

Before Transformers, most state-of-the-art sequence models were based on gated versions of RNNs, such as long short-term memory networks (LSTMs)~\cite{h4} and gated recurrent unit networks (GRUs)~\cite{h5}.
Compared to RNNs, Transformers can more easily capture long-term dependencies in sequences and facilitate efficient parallelization, enabling training on large datasets. These Transformer capabilities have led to the development of pre-trained language models such as BERT~\cite{h6} (Bidirectional Encoder Representations from Transformers) and GPT~\cite{h7} (Generative Pre-trained Transformer), trained with large text data sets. 

Quantum computers hold the promise of delivering distinct advantages over classical counterparts in specific machine-learning applications~\cite{h8}. By capitalizing on quantum mechanics principles, notably superposition and entanglement, quantum computers can execute computations in manners beyond the reach of classical computers. 
As part of investigating possible benefits of quantum computing for machine learning, researchers have been interested in exploring the possible application of quantum computing to sequence modeling by integrating quantum computing approaches with state-of-the-art classical sequence models like RNNs and Transformers.  This has led to the development of such quantum machine learning models as Quantum Recurrent Neural Networks (QRNN)~\cite{h22} and Quantum Self-Attention Neural Networks (QSANN)~\cite{h29}. These models aim to harness quantum properties to enhance sequential data processing and augment conventional attention mechanisms. Through the utilization of quantum entanglement and superposition, these models have the capability to represent and compute certain complex transformations more efficiently than with classical computing and models, which could, for example, be used in the transformation steps of RNN or Transformer models.  It is crucial to emphasize, however, that practical and scalable quantum computers are still in early stages of development. 

The paper first introduces quantum neural networks as the fundamental building blocks of the different models that follow, followed by describing different quantum data encoding methods used. The subsequent sections focus on a literature survey on quantum recurrent neural networks(QRNNS) and quantum self-attention neural networks (QSANNs) for text and image classification. 

As part of exploring this quantum sequence model work, we implemented a key representative set of these quantum algorithms and included experimental comparison across multiple datasets for each application type, comparing as well to classical counterparts to show their differences in performance.  We also explored a few basic enhancements for these quantum models, including exploring different encoding techniques and a parameterized ansatz for data loading, adapting an image classification approach using quantum self-attention to create a hybrid quantum transformer model that works for images and text, and introducing positional encoding into quantum self-attention, for which we observed improved accuracy and faster convergence in image and text classification experiments.
 The relevant code and results are publicly available on github~\cite{git}.

\section{Quantum Neural Networks}
Quantum neural networks (QNNs) capitalize on fundamental quantum principles like superposition and entanglement, which allow QNNs to represent multiple states simultaneously, or from a different view, represent a much larger state with exponentially fewer qubits than classical bits, potentially exponentially expanding their capacity for parallel processing. Entanglement enables quantum bits (qubits) to be correlated in ways that classical bits cannot, facilitating more complex and efficient computations. Some proposals for quantum neural networks include~\cite{h8,h9, h10, h11} and allude to possible benefits such as accelerated training and enhanced processing speed. QNNs are a specific category within variational quantum algorithms, consisting of quantum circuits that incorporate gate operations parameterized for optimization~\cite{h12}, which combines quantum computing principles with deep learning techniques~\cite{h13}. First, classical data is typically transformed into a quantum state using a quantum feature map to encode each data point~\cite{h14}. After encoding, a variational model is applied. This model comprises parameterized quantum gate operations that are subsequently optimized for a specific task, akin to classical machine learning approaches~\cite{h9, h10, h11}. 
 Finally, the output of the quantum neural network is then obtained by measuring the quantum circuit. These measurements are typically transformed into predictions through classical post-processing. For example, measurements on all qubits are done in the $\sigma{_z}$ basis and then the parity of the resulting bit strings is calculated classically to serve as a binary class prediction. To simplify, we focus on binary classification, where the probability of class 0 is associated with the probability of encountering even-parity bit strings, and likewise, for class 1, the probability of encountering odd-parity bit strings. Subsequently, these are fed into a loss function with the aim of selecting variational model parameters that minimize the loss, averaged across a training dataset, via  optimization such as gradient descent.  

\section{Quantum Data Encoding Techniques}
Quantum data encoding techniques are methods used to represent classical data as quantum states, allowing quantum computers to manipulate this information. Some common quantum data encoding techniques are: (1) Basis Encoding where quantum bits or qubits encode classical data in various bases, the most common being the computational basis, where $\ket{0}$ and $\ket{1}$ represent classical bits 0 and 1, respectively.  
(2) Amplitude Encoding ~\cite{h15}, where the probability amplitudes of a quantum state represent the values of classical data, enabling encoding vectors and matrices efficiently. 
For example, the state $\alpha \ket{0} + \beta \ket{1}$ can represent the classical vector [$\alpha, \beta$]. (3) The Angle Encoding method ~\cite{h16} uses one or more rotation operations, phase or otherwise, to encode each feature, where the rotation angle is derived from the feature, and is the most commonly used in variational QML methods due to its efficiency. \\
For QRNN, the data encoding circuit is as shown in fig.\ref{fig:Encoding}  where the dataset $x^{(t)} \in\left[0,1\right]^{\otimes n} $ at step $t$ is embedded into the quantum circuit via angle encoding $U_{in}$, with angles of the $R_Y$ gates via the inverse cosine function. The encoding of $i^{th}$ data can be expressed as 
    $R_{Y}\left(\theta_{x_i^{t}}\right)=\exp\left\{ i\theta_{x_i^{t}}Y_{i}\right\} $
where rotation angle $\theta_{x_i^{t}}=\cos^{-1}\left(x_{i}^{t}\right)$.
\begin{figure}
\begin{minipage}[t]{0.34\columnwidth}
  \includegraphics[width=\linewidth]{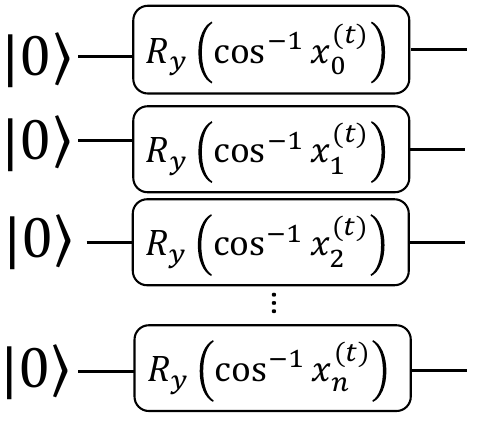}
  \caption{Data Encoding for QRNNs\label{fig:Encoding}}
\end{minipage}\hfill 
\begin{minipage}[t]{0.56\columnwidth}
  \includegraphics[width=\linewidth]{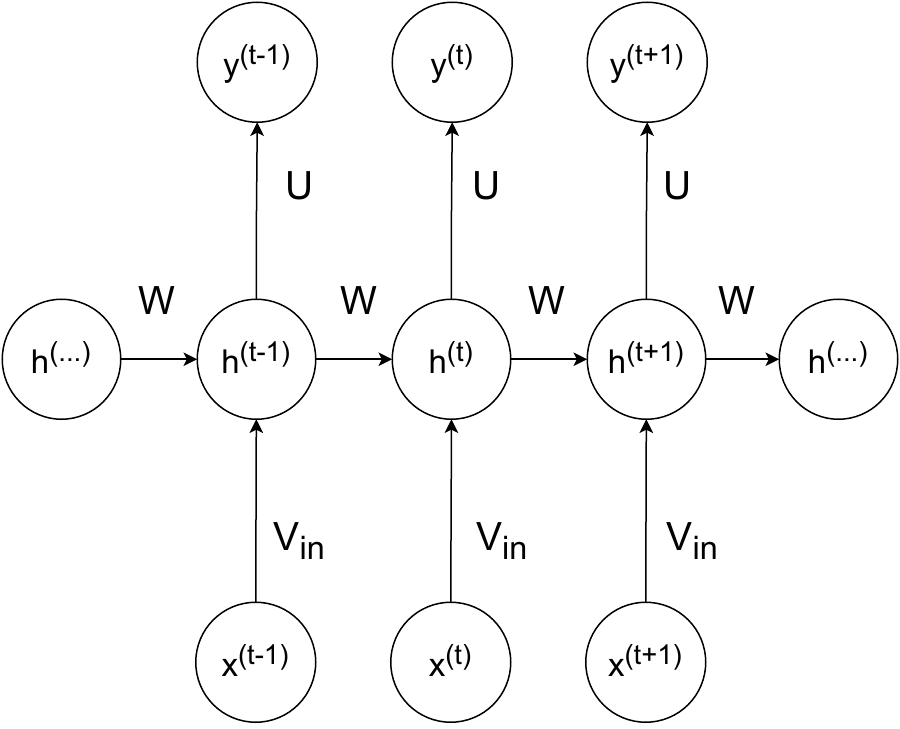}
  \caption{RNN state evolution\label{fig:rnnarch}}
\end{minipage}
\end{figure}

For QSANN, the data encoding circuit is shown in fig.\ref{fig:ansatz}. The classical input data ${\vec{x}_i\in \mathcal{R}^{d}}$ with $i=0,1,\cdots S$ is encoded into QSANN using a quantum feature map $U_{enc}(\vec{\theta})$  and then the initial state becomes 
$\ket{\psi(\vec{x}_i)} = U_{enc}(\vec{x}_i)  \ket{0^n}$.


After the classical data is mapped into quantum states via the data encoding, the QRNN and QSANN algorithm will further transform this state, as described in the next sections.

 \begin{figure}
     \includegraphics[width=0.45\textwidth]{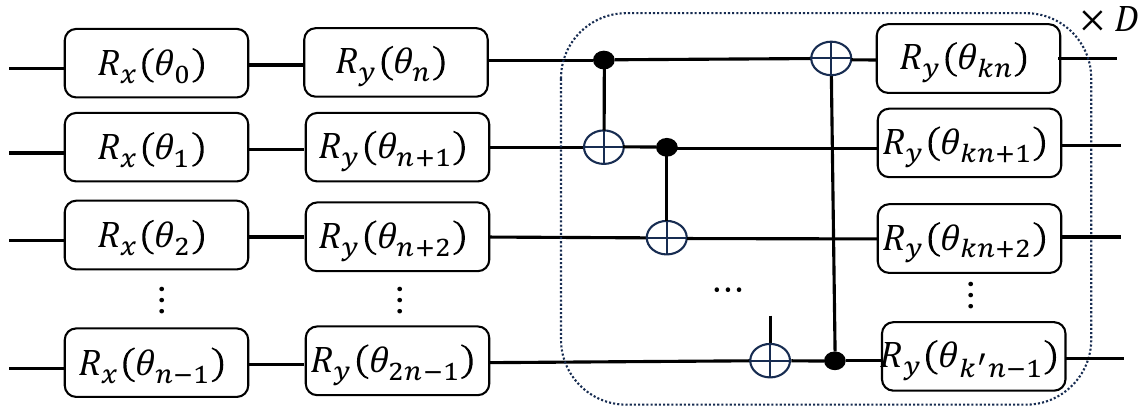}
     \caption{Data Encoding for QSANNs}
     \label{fig:ansatz}
 \end{figure}

\section{RNN for Sequence models}





\subsection{Classical Recurrent Neural Network}
The Recurrent Neural Network (RNN) ~\cite{h2, h17} is a natural generalization of the feedforward neural network to sequences. Given a sequence of inputs $(x_{1},\cdots, x_{n} )$, a standard RNN computes a sequence of outputs $(y_{1},\cdots, y_{n} )$ by iterating through the sequence. An RNN uses both present and past elements of a sequence to make predictions, where the past elements are encoded into an internal, or \emph{hidden}, state, as illustrated in Figure \ref{fig:rnnarch}, where $h^{(t)}$ is the hidden state, $x^{(t)}$ is the input, and $y^{(t)}$ is the output, at step $t$. 

RNNs consist of multiple blocks, and the history of each block is fed into the successive block. RNNs excel at mapping sequences to sequences with fixed patterns between input and output, but applying them to input and output sequences of varying or longer lengths with intricate and non-monotonic relationships generally remains a more challenging task.

\subsection{Quantum Recurrent Neural Network}
QRNNs are neural networks that utilize quantum circuits as a fundamental component. They serve as the quantum analogue of RNNs, utilizing quantum bits (qubits) and quantum gates to process sequential data and capture dependencies within it. QRNNs can be hybrid or purely quantum models. Hybrid QRNNs typically use quantum circuits for acceleration~\cite{h22}. In fully quantum QRNNs, the non-linear transformations of the recurrent blocks are replaced  entirely by quantum circuits, making them Quantum Recurrent Blocks (QRBs). QRNNs are composed of multiple such QRBs. The QRB, essentially a parameterized quantum circuit, is analogous to classical recurrent blocks in RNNs. QRBs require a data encoding function and an ansatz to encode and process classical data. The QRBs in general are dense, thus allowing parameter compression. The QRBs can also potentially avoid gradient decay as a result of unitary RNN structures ~\cite{h19}.
\\
A handful of QRNN algorithms have been proposed~\cite{h19,h20,h21}. ~\cite{h19} proposed one of the first QRNN models, consisting of a quantum neuron built using parameterized gates. This neuron in combination with amplitude amplification acts as a non-linear parameterized quantum neuron. ~\cite{h20} constructed a variational quantum recurrent neural network using parameterized quantum circuits. ~\cite{h21} builds on a quantum enhanced RNN based on continuous variable Quantum Neural networks, consisting of a layered architecture of parameterized quantum circuits. Here, we focus on one ~\cite{h22}, which is composed of many quantum recurrent blocks shown in fig\ref{fig:QRB}. The quantum recurrent block has two registers, one for input and output, and another register to store history and feed it into the ansatz. 
\begin{figure}[htbp]
    \centering
    \includegraphics[width=0.4\textwidth]{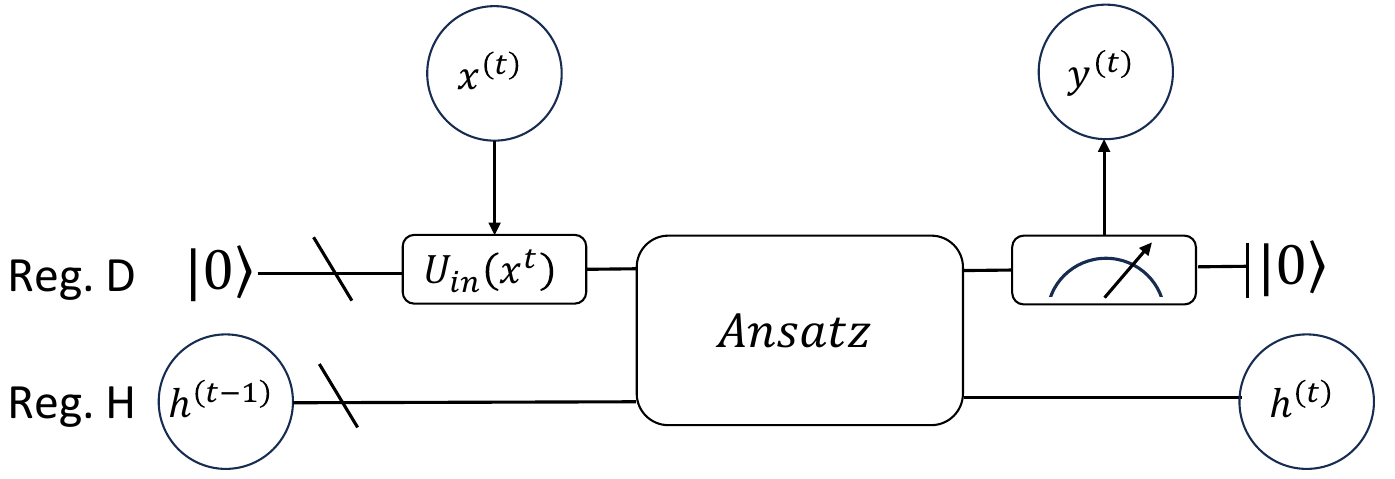}
    \caption{Quantum Recurrent Block}
    \label{fig:QRB}
\end{figure}

The ansatz in fig.\ref{fig:QRB} is a parameterized quantum circuit~\cite{h22}, where the rotation angles of quantum gates act as the learnable parameters of the neural networks. As shown in fig.\ref{fig:Ansatz}, the ansatz is composed of layers of two-qubit and single-qubit gates. The $X - Z$ decomposition was used for the single qubit gates in the circuit, $U_{1q} = R_x(\alpha)R_z(\beta)R_x(\gamma)$, where $\alpha$, $\beta$, $\gamma$ are respectively rotation angles of the $X$ , $Y$ and $Z$ gates and used as trainable parameters. Two-qubit $R_{zz}$ rotation gates were used to entangle qubits, $U_{2q} = R_{zz}(\theta) = e^{(i \theta Z_jZ_k)}$. 
In most quantum hardware devices, the $R_{ZZ}$ rotation gate is not an intrinsic gate. However, it can be implemented using two CNOT gates sandwiching an $R_Z$  gate. In fig.\ref{fig:Ansatz}, cyclic entanglement is used.
\begin{figure}[htbp]
    \centering
    \includegraphics[width=0.5\textwidth]{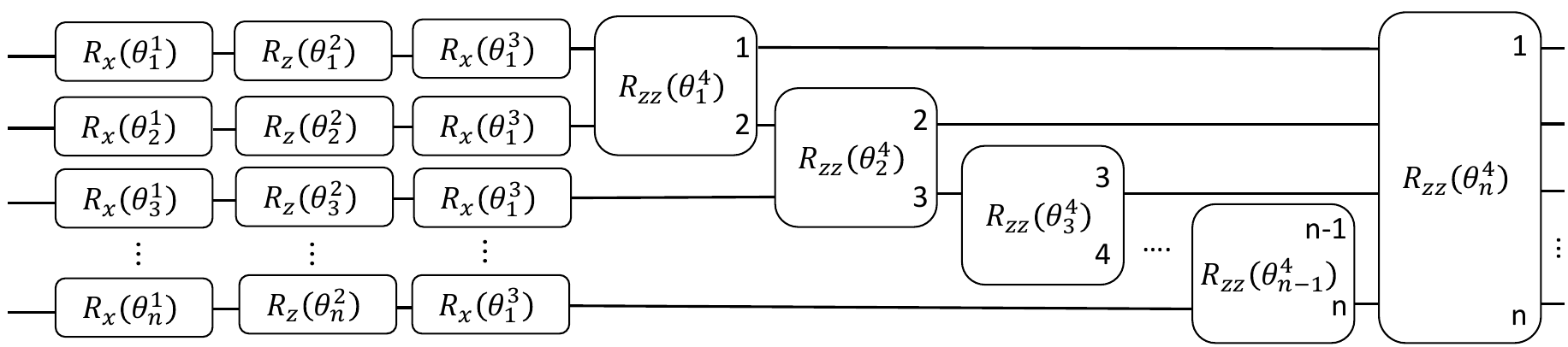}
    \caption{Ansatz Employed for QRNNs}
    \label{fig:Ansatz}
\end{figure}

After the ansatz, the first qubit from register D is measured using the Pauli $Z$ measurement, and the probability of getting state $|1\rangle$ is the output $y^{(t)}$ for all steps and is used for the prediction of real-time data. We then measure all qubits of register D and reset them to state $|0\rangle$.




To create a Quantum Recurrent Block(QRB) we combine encoding, ansatz, and the measurement circuits as shown in fig.~\ref{fig:QRB}. There are two ways to arrange the QRB, the plain architecture, called the plain QRNN (pQRNN), and the staggered architecture, called the staggered QRNN(sQRNN). In pQRNN, the same qubits are assigned to Reg. D and Reg H whereas in sQRNN they assign qubits to QRBs turn by turn. Thus the sQRNN needs less coherence time as compared to the pQRNN. Here the QRB block we employed uses pQRNN, which we chose due to the simplicity of implementation in the statevector simulator, since sQRNN requires dynamic circuits, applying the reset operation to reuse the QRBs qubits turn by turn.




For text data experiments, we tested the QRNN model (along with others), on the Meaning Classification (MC)~\cite{h24} and Relative Pronoun (RP)~\cite{h25} datasets, as well as the the Amazon, Yelp, and IMDB datasets~\cite{h26}. 
Here we employ the Term Frequency-Inverse Document Frequency (TF-IDF) method~\cite{h23} to vectorize the text data. TF-IDF is a text vectorization method which uses frequency to determine weights of words. 
We trained a QRNN on each dataset for 200 epochs and used a learning rate of 0.08. The results have been tabulated in Table.\ref{tab:qsann_m_qrnn}.

\section{Transformer architecture for sequence models}

\subsection{Classical Self-Attention}




The self-attention mechanism~\cite{h3} is employed in Transformer-based models, such as BERT and GPT, to capture contextual relationships within a sequence of data. This approach helps the model to identify dependencies and patterns by allowing each element in the input sequence to pay attention to and consider its relevance with respect to all other items. 
Each element, $\vec{x}_{i}$ is dynamically updated:  $\vec{x}^{\text{new}}_{i}=\sum_{j}w_{ij}\vec{v}_{j}$, where $\vec{v}_j$ is the \emph{value} for $\vec{x}_j$, and $w_{i,j}$ is the attention weight between element $i$ and $j$,
$w_{ij}=\text{Softmax}_j(\vec{q}_{i}\cdot\vec{k}_{j} / \sqrt{d})$.  Here $\vec{q}_i$ and $\vec{k}_j$ are \emph{query} and \emph{key} vectors respectively, and $\sqrt{d}$ is a scaling factor. These context vectors $\Vec{v}_i, \Vec{q}_j, \Vec{k}_h$ are given as $\vec{v}_i=U_{v}\vec{x}_i, \vec{q}_j=U_{q}\vec{x}_j$, and, $\vec{k}_h=U_{k}\vec{x}_h$ where $U_{i=k,q,v}$ are the conversion matrices, the elements of which are trainable parameters determined by training the model. Values, keys, and queries are packed together into matrices $V, Q, K$ resp.

 \begin{figure}
 \centering
     \includegraphics[width=0.45\textwidth]{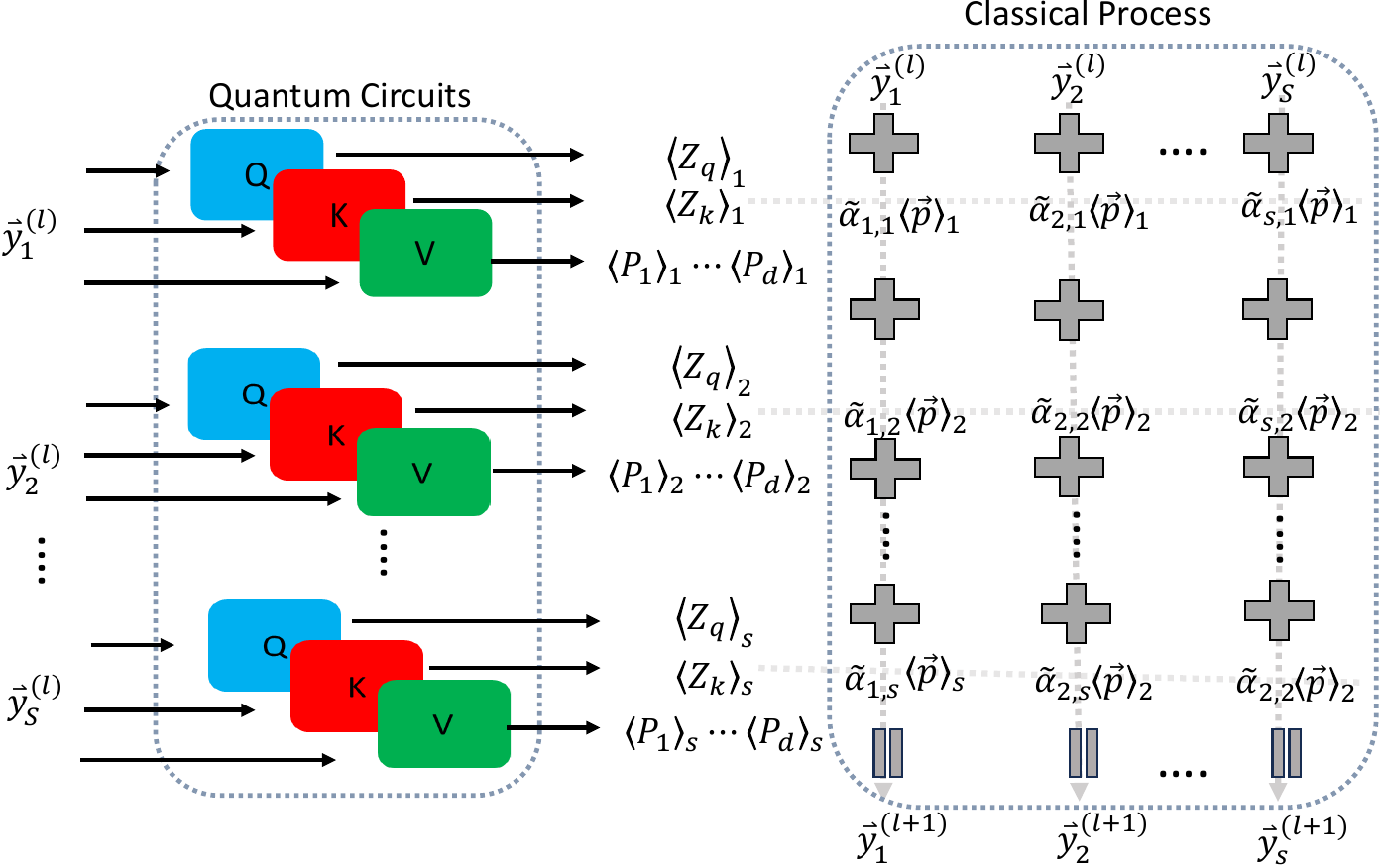}
     \caption{Quantum self-attention layer.}
     \label{fig:QSAL}
 \end{figure}

\subsection{Quantum Self-Attention Architecture}

There are two main quantum self-attention models that have been proposed so far. One can be realized by a full quantum circuit without classical post-process~\cite{h27}. It utilizes a quantum bit self-attention score matrix and quantum logic similarity as part of the quantum self-attention mechanism. However, this method requires plenty of qubits to be executed. Another model consists of several quantum variational circuits which use a Gaussian projected self-attention method~\cite{h28} and then post-processing using classical calculations. Here, because of the limitation of quantum simulators and quantum hardware in the near term, we focus on the latter. This quantum self-attention neural network (QSANN) is composed of many quantum self-attention layers (QSALs).  Each layer generates output $\vec{y}^{(l)}_i$, which is used as the input of the next QSAL. 

The details of a QSAL are highlighted in Fig.\ref{fig:QSAL}. QSAL consists of a quantum circuit part and a classical post-processing part. In the quantum part, for each sequence, there are three main different parameterized quantum circuits, $U_q$, $U_k$, $U_v$ which are employed to represent the query, key, and value components of the self-attention model. All use the same variational ansatz shown in Fig.~\ref{fig:ansatz} but with different variational parameters $\Vec{\theta}_q$, $\Vec{\theta}_k$, and $\Vec{\theta}_v$. The outputs of the query and key parts are the expectation value $\langle Z_0 \rangle$ of the first qubit from the parameterized quantum circuit, i.e., $\langle Z_j \rangle = \langle \psi (\vec{y}^{(l)}_i) | U^{\dagger}_{j}(\theta_j)Z_0 U_{j}(\theta_j) | \psi (\vec{y}^{(l)}_i) \rangle
$, where $j=q$ for the query and $j=k$ for the key,
for input $\vec{y}^{(l)}_i$.
The value output is made up of $d$ expectation values of $d$ Pauli string observables $ P_j \in (I, X, Y, Z)^{\otimes n} $ and forms a $d$-dimensional vector,
   $\Vec{v}_s = [\langle{P_1}_s\rangle, \langle{P_2}_s\rangle, \cdots, \langle{P_d}_s\rangle]^T $, where,
$ \langle{P_j}\rangle = \langle{\psi (\vec{y}^{(l)}_i)}| U^{\dagger}_{v}(\theta_v)P_j U_{v}(\theta_v) |{\psi (\vec{y}^{(l)}_i)}\rangle$.
In the classical part shown in the right of Fig.\ref{fig:QSAL}, the measurement output from the key and query's quantum parameterized circuits are used to calculate their similarity using Gaussian projected self-attention, $\alpha_{s,j} = e^{-(\langle Z_q \rangle_s - \langle Z_k \rangle_j)^2}$,
where $\alpha_{s,j}$ is the gaussian projected self attention coefficient between $s$ th input vector and $j$ th input vector. With this coefficient, the output of QSAL can be computed as 
\begin{equation}
    \vec{y}_{i}^{(l)}=\vec{y}_{i}^{(l-1)}+\sum_{j=1}^{S}\tilde{\alpha}_{s,j}\vec{v}_{j}
\end{equation}
where $\tilde{\alpha}_{s,j}$ is the normalized gaussian projected self attention coefficient $\tilde{\alpha}_{s,j}= \alpha_{s,j} / \sum_{m=1}^{S}\alpha_{s,m}$.

\begin{figure}[htbp]
  \centering
  \includegraphics[width=0.45\textwidth]{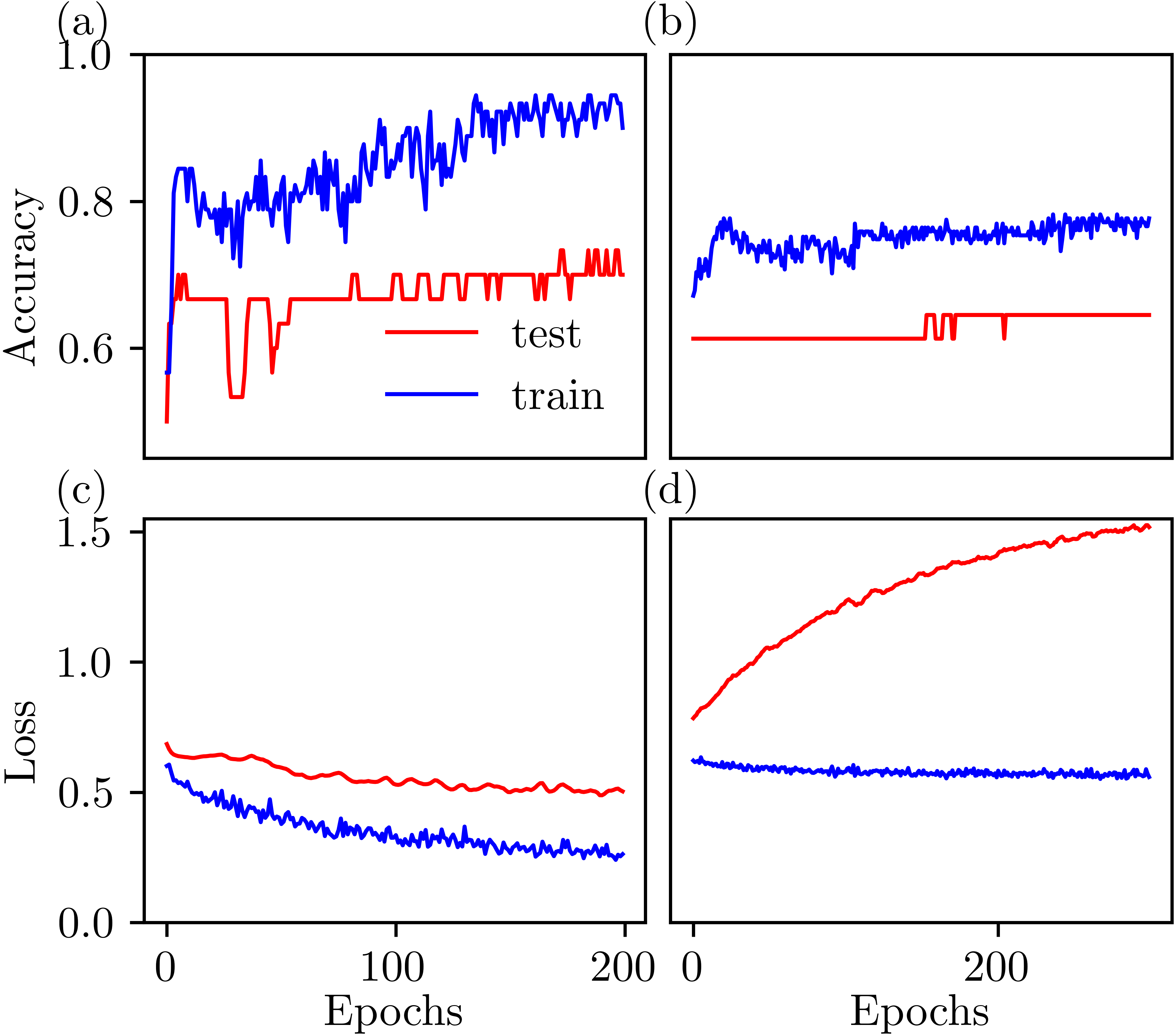}
  \caption{QSANN result for the MC and RP task without positional encoding. (a) The accuracy and (c) loss function for MC task (left-hand side column). (b) The accuracy and (d) loss function for the RP task (right-hand side column).}
  \label{fig:WO_pos_enco}
\end{figure}

\begin{figure}[htbp]
  \centering
  \includegraphics[width=0.48\textwidth]{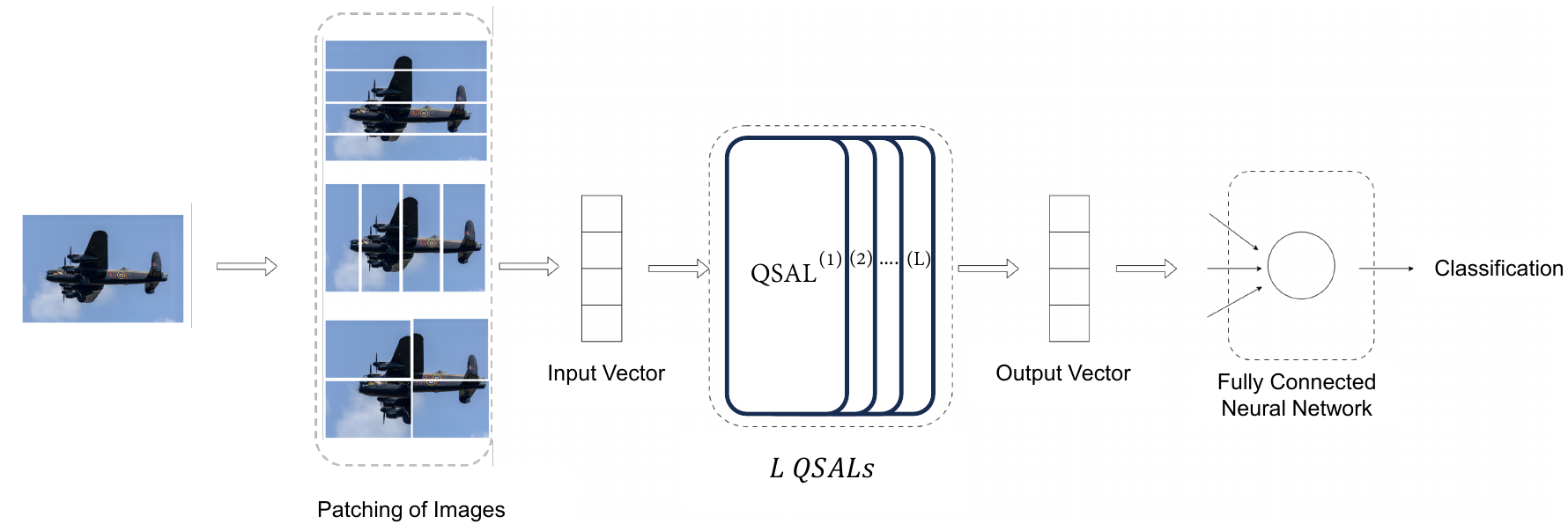}
  \caption{Image Classification with QSANNs}
  \label{fig:figqsann}
\end{figure}

\begin{table*}[h!]
\begin{tabular}{ c|cccccccccc }
\hline
Dataset & No. of Samples & Epoch & Train  & Test & Train & Test & Train &  Test & Train & Test\\ & & & Acc.(QSANN) & Acc. (QSANN) &  Acc.(CT) & Acc.(CT) & Acc(QRNN). & Acc(QRNN). & Acc(RNN). & Acc(RNN).\\
\hline
MC & 130 & 180 & 1 & 1 & 1 & 1 & 0.986 & 0.967 & 1 & 1\\
RP & 105  & 300 & 1 & 0.93 & 0.99 & 0.59 & 0.62 & 0.613 & 1 & 1\\
Amazon & 1000 & 300 & 0.93 & 0.64 & 0.94 & 0.67 & 0.653 & 0.658 & 1 & 1\\
IMDB & 1000 & 300 & 0.89 &  0.57 & 0.98 & 0.55 & 0.554 & 0.859 & 1 & 1\\
Yelp & 1000 & 300 & 0.95 & 0.72 & 0.97 & 0.6 & 0.695 & 0.602 & 1 & 1\\
\end{tabular}
\centering
\caption{The result of text Classification using QSANN, QRNN, RNN and classical transformer (CT) from the paper~\cite{h31}.}
\label{tab:qsann_m_qrnn}
\end{table*}
\subsection{Quantum Self-Attention for Text Classification}

Text Classification can be done by dividing the text into several pieces and feeding them into the QSANN. This also requires the standard practice of transforming text into a vector, done here based on the rank of how often the word shows up in the training dataset. For the initial execution of the model, we use two simple datasets, the MC task, and the RP task, to test our methods. MC task has 70 training samples (containing 17 words) and 30 test samples. RP task has 74 training samples (containing 91 words) and 31 test samples. Each sample includes a sentence with four words.  Since each QSAL accepts $n\times (D_{\text{enc}}+2)$ features, we could extend the vector to a 6-length vector by adding two zeros after vectorizing the sentence. We set $D=2$, one layer of QSAL, and the learning rate $0.005$ for both tasks. Fig.~\ref{fig:WO_pos_enco} shows the results. After 200 epochs, QSANN achieves $73$\% test accuracy on MC task *Fig.~\ref{fig:WO_pos_enco}(a)). However, for the RP task QSANN gets $64.5$\% accuracy (Fig.~\ref{fig:WO_pos_enco}(b)). The early growth of the test loss for the RP task shown in Fig.~\ref{fig:WO_pos_enco}(c) indicates overfitting. The RP task is to identify a noun as a subject or an object, which is highly dependent on the word positions; thus, QSANN without positional information, as originally proposed, cannot predict the RP task well.

\subsection{Quantum Self-Attention for Image Classification}
Image classification can be done with quantum self-attention neural networks in a similar way to text classification, as described in Fig.~\ref{fig:figqsann}. The first step is to divide an image into smaller patches so that each patch acts like a sequence in the text classification model. The patching can be done in three different ways: row-wise, column-wise, and block-wise. We found different type of patching did not largely influence the performance of the model in our case for the chosen datasets, so we preset results for row-wise patching herein.  The patched image is then embedded into the quantum circuit. This step is simpler than the text classification case as the image data is already in the form of numerical values (pixel values). As typical, image data is scaled to a $[0, 1]$ range.
The input vectors are then passed through multiple layers of quantum self-attention layers (QSALs), and the output from the QSALs is then fed to a fully connected neural network for classification.

\begin{table*}[t!]
\begin{tabular*}{\linewidth}{@{\extracolsep{\fill}} c|ccccccl}
\hline
Dataset& \# Classes & \# Samples & \#Epochs & Train Acc. & Test Acc. & Train Acc. & Test Acc. \\
 & & & & (CVT) & (CVT) & (QSANN) & (QSANN)\\
\hline
Sk Digit & 2 &  280 &  10 & 1.0 & 1.0 & 1.0 & 0.98 \\
Sk Digit & 10  & 100 &  10 & 1.0 & 0.98 & 1.0 & 0.9 \\
Sk Digit & 10  & 500 &  30 & 1.0 & 0.95 & 1.0 & 0.84 \\
MNIST & 10  & 500 &  30 & 0.99 & 0.96 & 0.98 &  0.85 \\
MNIST & 10  & 1000 &  50 & 0.96 & 0.94 & 1.0 & 0.78 \\
FashionMNIST & 10  & 500 &  50 & 0.97 & 0.96 & 0.88 & 0.65\\
\end{tabular*}
\centering
\caption{The classification accuracy from QSANN and classical vision transformer (CVT). Here, results show the performance of QSANN without positional encoding as compared to CVT. The learning rate was also kept fixed throughout different datasets at 0.001. Out of all images, as highlighted in the table, 75\% were used for the training and 25\% for testing.}
\label{tab:freq}
\end{table*}

\subsection{Positional Encoding in QSANNs}

The QSANN method used for image and text classification so far does not employ positional encoding techniques to leverage spatial information of the embedded data. We found that the model converges faster for this image data when positional encoding was added, as shown in Fig~\ref{fig:figqsannpos}.

\begin{figure}[htbp]
  \centering
  \includegraphics[width=0.35\textwidth]{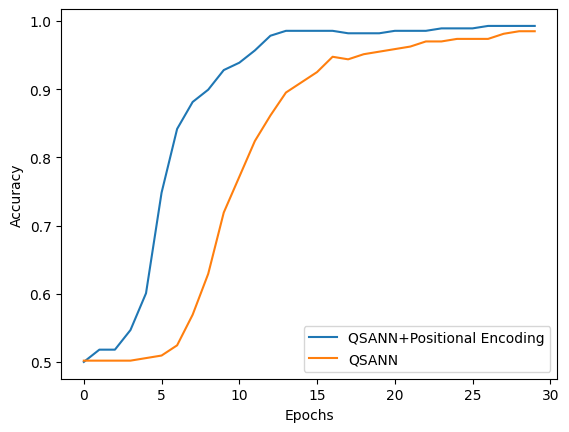}
  \caption{Comparison of the performance of QSANNs for Image Classification with and without positional encoding (dataset is Sklearn's Digits dataset).}
  \label{fig:figqsannpos}
\end{figure}

\begin{figure}[htbp]
  \centering
  \includegraphics[width=0.37\textwidth]{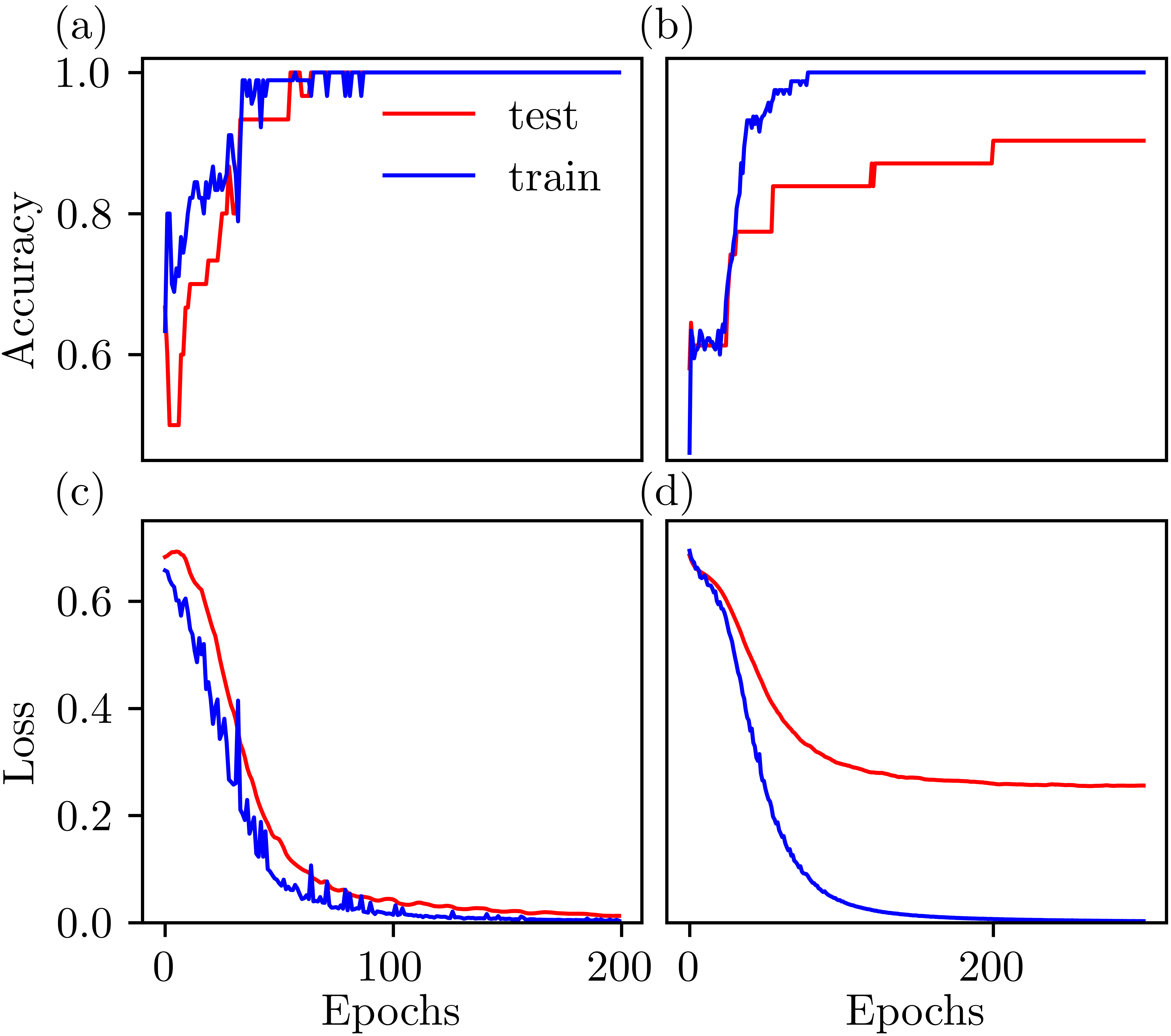}
  \caption{QSANN result for the MC task and RP task with positional encoding. (a) The accuracy and (b) loss function for MC task (left-hand side column). (c) The accuracy and (d) loss function for the RP task (right-hand side column).}
  \label{fig:W_pos_enco}
\end{figure}

To improve performance for both text tasks, we added an embedding layer and positional encoding layer to the QSANN. We set the embedding dimension as $2$ and the number of pieces that the text is divided as 2. We keep $D=2$, $n_q = 2$, and learning rate $0.005$. Fig.~\ref{fig:W_pos_enco} shows that positional encoding dramatically improves QSANN's . With positional encoding, QSANN can achieve $100$\% train and test accuracy for the MC task and $100$\% ptrain and $90.32$\% test accuracy for the RP task. Fig.~\ref{fig:W_pos_enco} (c) and (d) show that the training and test losses converge faster with positional encoding, even though there are more trainable parameters.   

\section{Results and Conclusions}
This work highlights a small portion of the ongoing work on quantum sequential models, including QRNNs and quantum self-attention models, showing the applicability and versatility of QSANNs across text and image classification tasks. QSANN, enhanced by positional encoding, achieves superior performance by leveraging its self-attention mechanism to emphasize crucial words or essential elements within an image or text sequence. This mechanism allows the model to discern and prioritize important information effectively. In contrast, QRNN faces challenges in capturing long-term dependencies due to its inherent short-term memory limitations. 
Classical self-attention excels in a variety of tasks, including text summarization and machine translation, by giving an adaptive and contextualized representation of the input material. 
It is a crucial innovation in contemporary deep learning systems because of its parallelization and versatility. 
Quantum self-attention holds the potential to further leverage and expand these properties, and transform  machine learning applications, by potentially enabling faster and more accurate solutions to intricate modeling tasks.

\section*{Acknowledgment}
We acknowledge libraries Qiskit~\cite{h33},  PennyLane ~\cite{h34} and Pytorch ~\cite{h35}, and the Qiskit Advocate Mentorship Program (QAMP) in which we conceptualized and developed this work.


\end{document}